# Magnetic Reconnection in the Plasma Disk at 23 Jupiter Radii

Jian-zhao Wang[1,2], Fran Bagenal[1], Stefan Eriksson[1], Robert E. Ergun[1,2], Peter A. Delamere[3], Robert J. Wilson[1], Robert W. Ebert[4,5], Philip W. Valek[4], Frederic Allegrini[4,5], Licia C. Ray[6]

[1] Laboratory for Atmospheric and Space Physics, University of Colorado Boulder, Boulder, CO, USA
[2] Department of Astrophysical and Planetary Sciences, University of Colorado Boulder, Boulder, CO, USA
[3] Geophysical Institute, University of Alaska Fairbanks, Fairbanks, AK, USA
[4] Southwest Research Institute, San Antonio, TX, USA
[5] Department of Physics and Astronomy, University of Texas at San Antonio, San Antonio, TX, USA
[6] Physics Department, Lancaster University, Lancaster, UK

*Corresponding author: J.-z. Wang (jiwa1124@colorado.edu; jian-zhao.wang@lasp.colorado.edu)

**A key open question in astrophysics is how plasma is transported within strongly magnetized, rapidly rotating systems. Magnetic reconnection and flux tube interchange are possible mechanisms, with Jupiter serving as the best local analog for distant systems. However, magnetic reconnection at Jupiter remains poorly understood. A key indicator of active magnetic reconnection is the ion diffusion region, but its detection at Jupiter has remained unconfirmed. Here, we report a unique magnetic reconnection event in Jupiter's inner magnetosphere that presents the first detection of an ion diffusion region. We provide evidence that this event involves localized flux tube interchange motion driven by centrifugal forces, which occurs inside a thin current sheet formed by the collision and twisting of two distinct flux tubes. This study provides new insights into Io-genic plasma transport at Jupiter and the unique role of magnetic reconnection in rapidly rotating systems, two key unresolved questions.**

Centrifugal forces play a significant role in structuring the Jovian magnetosphere, which is characterized by its strong intrinsic magnetic field, an internal plasma source from Io's escaping atmosphere, and fast planetary rotation (Achilleos et al., 2015). As a crucial aspect of the system's dynamics, magnetic reconnection is a key mechanism in governing plasma behavior. At Jupiter, magnetic reconnection features a dominance of heavy ions, possible corotating reconnection sites (Zhao et al., 2024), and radial stress driven by centrifugal forces (Louarn et al., 2015), making it distinct from reconnection at Earth or other weakly magnetized planets. In the outer region beyond 30 $R_J$ ($R_J$=71,492 km is Jupiter's equatorial radius), weakened magnetic fields are stretched as plasma moves outward due to centrifugal forces. Ultimately, magnetic reconnection may occur at an intersection of opposing magnetic field lines, leading to the pinching off of plasmoids, as part of the Vasyliūnas cycle (Vasyliūnas, 1983).

The Io-genic plasma is transported outward and confined near the equator under centrifugal forcing, forming a plasma disk (Bagenal & Delamere, 2011). In the inner region within 30 $R_J$, radial plasma transport is facilitated by flux tube interchange, which occurs spontaneously when the flux tube content decreases radially (Southwood & Kivelson, 1987; Kivelson & Southwood, 2005). Due to the difficulty of fully interchanging whole magnetic flux tubes through ionospheric convection, it is proposed that plasma convection from localized interchange at the equator twists the magnetic field, forming a pair of off-equator reconnection sites. This process could transport Io-genic plasma radially outward by exchanging a portion of the flux tube, known as the localized flux tube interchange driven magnetic reconnection (X. Ma et al., 2016), although there is no observational evidence to date to support this mechanism.

Magnetic reconnection in the Jovian system is typically proposed from its observable byproducts, e.g., flux ropes, plasmoids, and dipolarization fronts (Vogt et al., 2010, 2014; Artemyev et al., 2020; Blöcker et al., 2023). Near the magnetic X-line of reconnection, ions decouple from the magnetic field, introducing an ion diffusion region with the presence of a quadrupole Hall magnetic field, which is a key indicator of local reconnection (Øieroset et al., 2001; Zhang et al., 2012; Arridge et al., 2016; Guo et al., 2018). The Hall field is generated as electrons stream along the separatrix regions toward the X-line and jet away from the X-line in two reconnection outflow regions. The ion diffusion region has remained undetected at Jupiter due to its small scale and absence of high-quality plasma data. The polar-orbiting Juno spacecraft around Jupiter enables the first high-cadence off-equator measurements to address whether flux tube interchange reconnection can proceed close to the planet as theorized. In this study, we present the first ion diffusion region detection at Jupiter in support of a flux tube interchange driven reconnection that facilitates outward Io-genic plasma motion inside 30 $R_J$.

## Overview of the Magnetic Reconnection Event

An overview of the event is shown in **Figure 1**. The magnetic field is measured by the MAG instrument (Connerney et al., 2017), and the plasma parameters are obtained using a forward modeling method (J. Wang et al., 2024a, 2024b) based on thermal plasma data from the JADE instrument (McComas et al., 2017). All observations are presented in the Jupiter-De-Spun-Sun (JSS) coordinate system. In spherical coordinates, near the equator, the $r$, $\theta$, and $\phi$ components approximate to the radial direction, the north-to-south direction, and Jupiter's rotational direction, respectively. As illustrated in panels (a) and (b), this event occurred at a radial distance of 23.3 $R_J$ and a local time of 22.6 hr, near the midnight magnetic equator. The event lies within the plasma disk, far from both the magnetopause (Joy et al., 2002) and the statistical magnetic reconnection X-line driven by the Vasyliūnas cycle, as identified by Galileo observations (Vogt et al., 2010). This location within the inner magnetosphere rules out mechanisms involving magnetosheath reconnection or reconnection driven by the Vasyliūnas cycle.

In panel (c), $B_r$ shifts from negative to positive, indicating that Juno was crossing the plasma disk at 01:55 UT, where $B_r \sim 0$, from south to north. Around this crossing, typical plasma disk features are observed, including a strong, stable $B_\theta$ and a plasma flow reaching 80% of the rigid corotational velocity ($v_\phi \approx 235$ km/s). South of the equator, Juno detects a structure with entangled flux tubes, characterized by its helical magnetic field (panel (d)) and a sharp density contrast between the two tubes (panel (e)). The first half of the structure features a tenuous, hot flux tube with increased $B_\theta$ that indicates inward stretching, while the second half contains a dense, cold flux tube with decreased $B_\theta$ that indicates outward stretching. The opposite trend in $B_\phi$ indicates that the two tubes are twisted around each other.

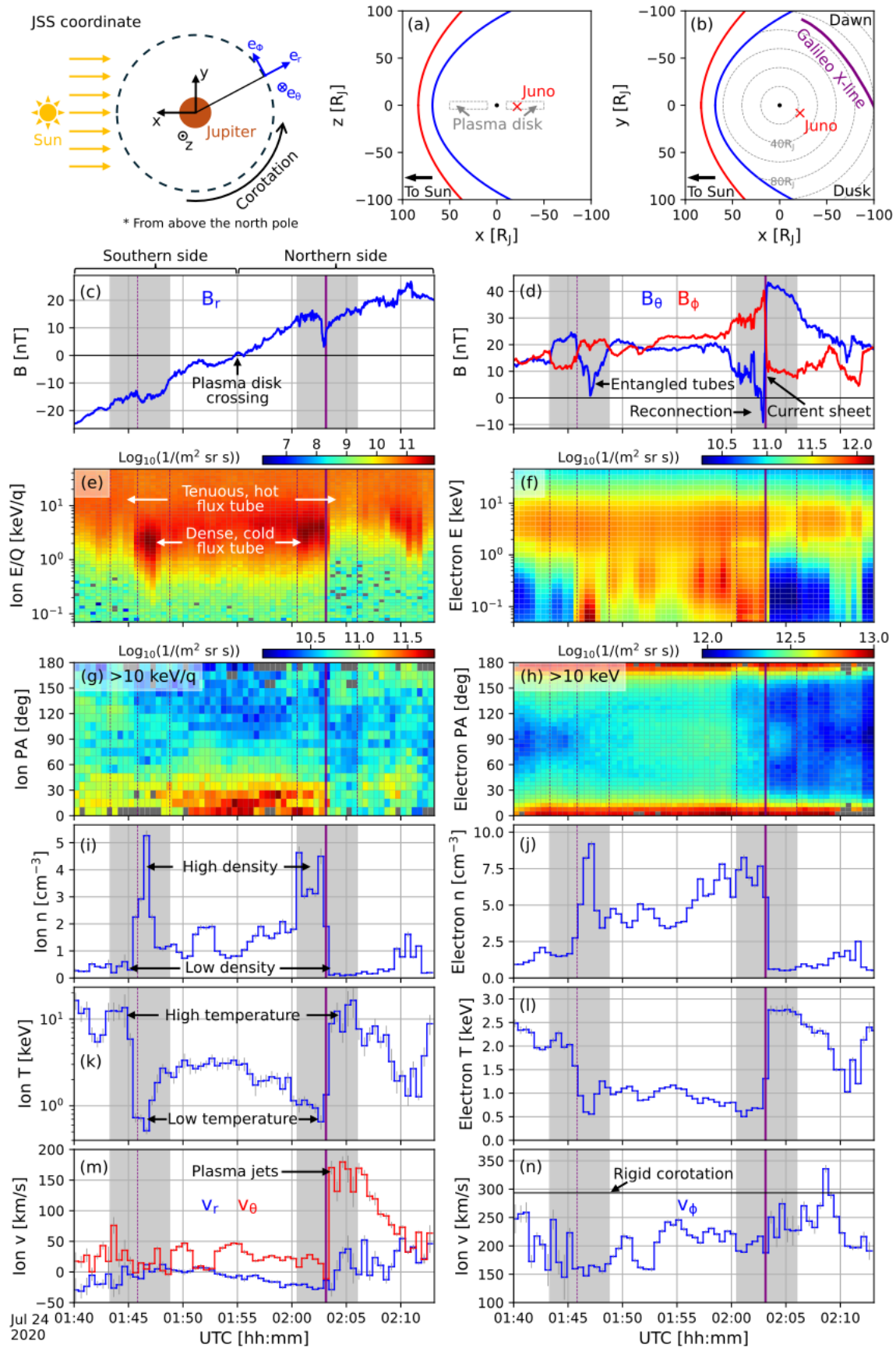

**Figure 1.** Overview of the magnetic reconnection event, including (**a,b**) the event location, (**c,d**) magnetic field measurements, (**e,f**) heavy ion and electron flux spectra, (**g,h**) pitch angle (PA) distributions, (**i,j**) plasma densities, (**k,l**) plasma temperatures, and (**m,n**) plasma velocities. All observations are presented in the Jupiter-De-Spun-Sun (JSS) coordinate system. The upper left panel shows an illustration of the JSS system near the equator. The gray-shaded regions mark the entangled flux tubes structure and the magnetic reconnection event. The thick, solid purple line indicates the location of the current sheet.

On the north side of the equator, Juno detects evidence of a magnetic reconnection event. First, $B_\theta$ increases sharply after a brief negative dip, rising from -9 nT to 43 nT in 20 s (panel (d)). This increase is 4.4 times greater than the average rise amplitude and 76% shorter than the median rise time of dipolarization fronts reported by Blöcker et al. (2023). Simultaneously, $B_\phi$ decreases from 40 nT to 10 nT. The rapid changes in both $B_\phi$ and $B_\theta$ suggest the presence of an intense and very thin current sheet where reconnection could potentially occur. The shear angle between the magnetic fields on opposite sides of the current sheet is 86.2 deg, which is sufficient to trigger magnetic reconnection, as observed at Jupiter (e.g., Montgomery et al. (2022)) and Earth (e.g., Kacem et al. (2018)). Second, both electron and ion data show a similar trend, with a dense, cold flux tube and a tenuous, hot flux tube closely located on opposite sides of the current sheet (panels (i)-(l)). Consistent with the southern structure, $B_\theta$ increases in the tenuous tube and decreases in the dense tube, but with greater variation, indicating that the two tubes are compressed against each other. This suggests that magnetic reconnection could occur within the thin current sheet formed at the center of two compressed, interlaced flux tubes (discussed further below). Plasma flow is also accelerated in the $\theta$ direction by ~100 km/s (panel (m)). The Alfvén speed is calculated to be 91 km/s, which closely matches the increase in flow velocity, showing consistency between the observations and the Walén relation (Sonnerup et al., 2018). Additionally, the input magnetic energy density of $6.7\times10^{-10}$ J/m$^3$ aligns with a plasma kinetic energy density gain of $8.0\times10^{-10}$ J/m$^3$, demonstrating consistency with the principle of energy conservation in magnetic reconnection. Third, pitch angle distributions (PADs) reveal that while electrons remain field-aligned (panel (h)), ions exhibit an asymmetric PAD (panel (g)), suggesting decoupling between ions and electrons. In summary, the observations of magnetic fields, plasma bulk flow parameters, and PADs are consistent with the detection of a reconnection event.

Considering the southern entangled flux tubes structure and the northern reconnection event together, these observations suggest a localized interchange driven reconnection process, supported by multiple lines of evidence. First, this event occurred at a radial distance of 23.3 R$_J$, which is near the outer boundary of observed interchange events (Daly et al., 2024) but significantly closer to Jupiter than the expected locations of Vasyliūnas cycle driven reconnection events (Vogt et al., 2020). Unlike whole flux tube interchange, partial and localized interchange is expected at this location due to the weakened magnetic field (X. Ma et al., 2019). Second, the two regions are symmetrically located with respect to the equator, where each interval of decreased $B_\theta$ contains a dense, cold flux tube, and each interval of increased $B_\theta$ contains a tenuous, hot flux tube. Quantitatively, the entangled tubes structure and reconnection event are located 0.39 R$_J$ south and 0.33 R$_J$ north of the plasma disk center, respectively. The observations also suggest the twisting of two flux tubes related to interchange, with the two highly correlated regions representing different stages of the process and observed by Juno at different times. The entangled tubes structure represents the early growth stage of twisting, where the magnetic field begins to distort. Approximately 15 minutes later, as Juno traveled from the southern to the northern disk, it encountered the correlated magnetic reconnection site. Here, the field was strongly twisted, and magnetic reconnection was triggered by a sufficient shear angle. Third, the two regions exhibit opposite polarities in magnetic field and plasma parameters, consistent with localized interchange driven reconnection, which shears the magnetic fields of the paired sites in opposite directions. In the southern region, the dip in $B_\theta$, along with a dense, cold flux tube, appeared in the second half. A similar signature was also observed in the northern region, but it occurred before the current sheet. The low $B_\theta$ magnitude could result from the radially stretched magnetic field during the mass-loaded flux tube's transport. Lastly, the reconnection outflow is directed in the $\theta$ direction (discussed further below), contrasting with typical Vasyliūnas cycle driven reconnection signatures or plasma injection events (Mauk et al., 1999; Louarn et al., 2015), where planetward jets are expected.

**Diffusion Region Observations**

**Figure 2**, panel (a), illustrates the mechanism triggering the magnetic reconnection event. Due to interchange motion driven by centrifugal forces, two interlaced flux tubes with distinct properties exert pressure on each other. This compression causes magnetic field pileup, leading to the formation of a thin current sheet between them. Consequently, the parallel components of the magnetic field in the two flux tubes correspond to a guide field, while the anti-parallel components facilitate magnetic reconnection, producing large-scale Alfvénic ion outflow. Juno traverses the current sheet from the high-density side to the low-density side, consistent with an asymmetric reconnection event characterized by a magnetic field shear angle of 86 deg and a density gradient. Similar reconnection events in thin current sheets have been observed in Earth's magnetosphere (Øieroset et al., 2016; Burch & Phan 2016; Kacem et al., 2018; Alexandrova et al., 2016; Qi et al., 2020).

To investigate the diffusion region near the thin current sheet, the observations are re-organized in the local current sheet LMN coordinate system (Eriksson et al., 2024; R. Wang et al., 2024). As illustrated in panel (b), near the X-line of magnetic reconnection, the *N*-direction is normal to the current sheet, the *M*-direction aligns with the Hall magnetic field, and the *L*-direction represents the outflow direction. As shown in panel (c), the $B_L$ component changes from positive to negative, reflecting the sharp magnetic field rotation as Juno traverses from one side of the thin current sheet to the other. The Hall quadrupolar magnetic field, identified by the bipolar reversal in $B_M$, provides direct evidence of the detected diffusion region, while a background of 33 nT in the $B_M$ profile suggests a substantial guide field. The $B_M$ pattern is 'shunted away' from the current sheet center (where $B_L$ is zero), consistent with the guide field asymmetry effect on the Hall field (Eastwood et al., 2010).

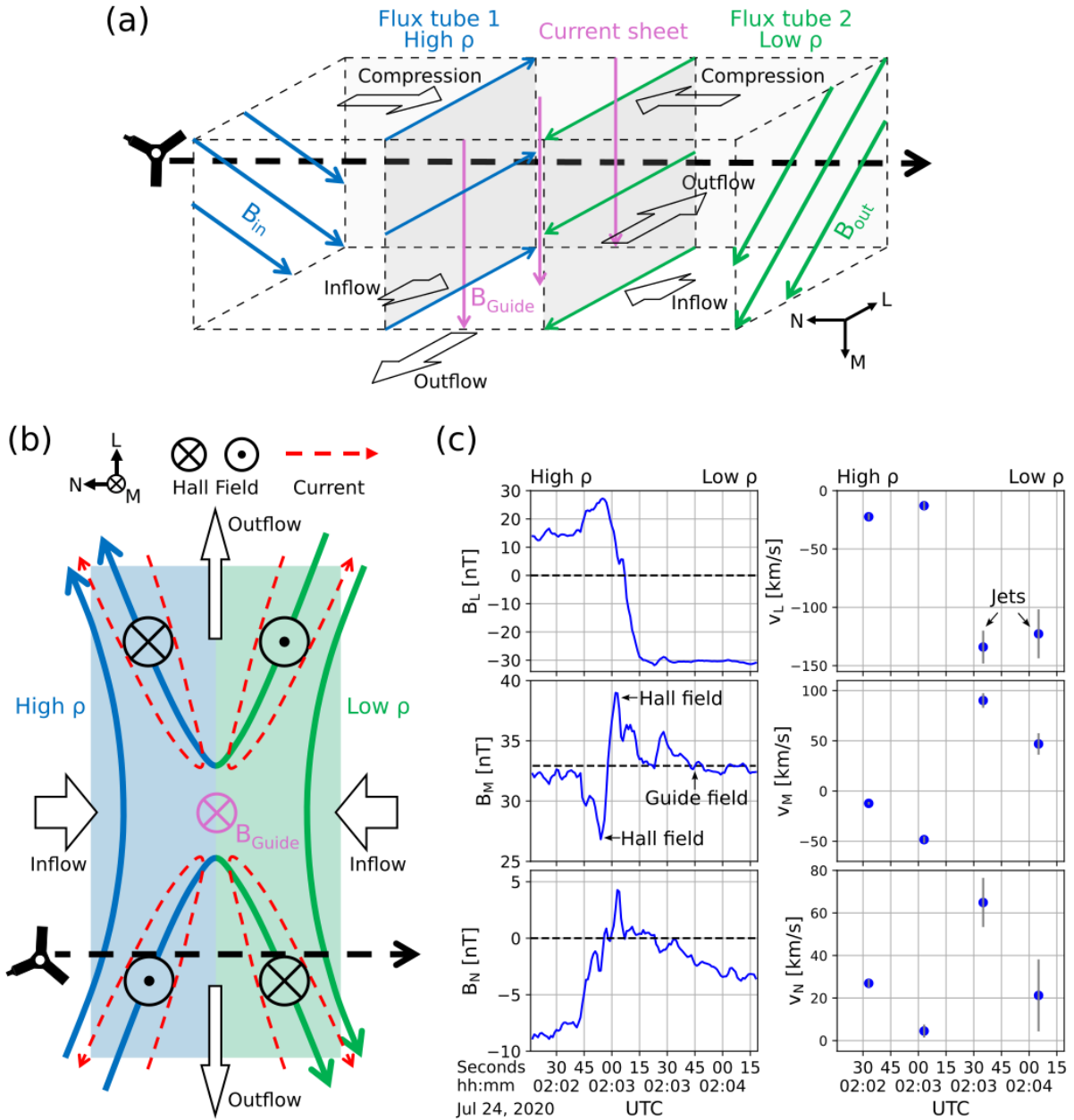

**Figure 2.** Overview of observations in the ion diffusion region (IDR). (**a**) Illustration of the magnetic reconnection event trigger mechanism. (**b**) Illustration of the diffusion region and Juno's trajectory. (**c**) Magnetic fields and plasma flows in the LMN coordinate system.

The plasma flow velocities in the LMN system are presented in the co-rotating frame (panel (c)), converted from velocities in the JSS system under the assumption that the reconnection site co-rotates with the planet. Plasma jets resulting from magnetic reconnection are observed in the low-density flux tube region, with accelerated velocities primarily in the negative $L$-direction, consistent with the proposed trajectory. The maximum $v_L$ outflow is offset from the current sheet center toward the low-density side, which is expected in reconnection with high density asymmetry. Plasma flow is also accelerated in the $M$-direction, consistent with the influence of the 33 nT guide field. This guide field modulates the outflow jets, causing them to align parallel to its direction (Pucci et al., 2018; Pritchett, 2015). Quantitatively, the ion inertial length, $d_i = C/\omega_{pi}$, is estimated to be 370 km. Juno traverses the current sheet in 25 s, defined as the time interval between the positive and negative maxima of $B_L$. Neglecting Juno's motion and assuming a flow along the normal to the current sheet at an average $v_N$~29 km/s (panel (c)), this corresponds to a normal width of 725 km or ~2 $d_i$, which is generally consistent with the estimated ion inertial length.

**Figure 3**, panel (a), shows the energy-dependent PADs of thermal electrons from JADE and energetic electrons from JEDI (Mauk et al., 2017). In addition to data from the dense tube and low-density tube regions of the magnetic reconnection event, observations from a separate plasma disk crossing at 21 $R_J$ are included for comparison as representative of typical background conditions without significant perturbation. This crossing occurred about four hours after the reconnection event on the same day. Panel (b) shows the PADs at three selected energy levels. In the background plasma, electrons exhibit field-aligned distributions across all energies, consistent with typical regions outside 15 $R_J$ in the absence of magnetic reconnection (Q. Ma et al., 2021). In contrast, the dense tube region shows a broader energy spectrum, while electron fluxes below 1 keV in the low-density tube are significantly depleted after the magnetic reconnection. Additionally, in the low-density tube region, electron fluxes are notably enhanced near 90 deg pitch angles for energies above 10 keV. Perpendicular electron energization is often observed in association with magnetic reconnection near the X-line (Xiong et al., 2022; Oka et al., 2023; C. Wang et al., 2024). Panel (c) shows the combined JADE and JEDI energy spectra for both regions, summed over all pitch angle bins. Differences between JADE and JEDI fluxes are reconciled to produce smooth spectra. The average spectrum at 25 $R_J$ from Liu et al. (2024) is also shown for comparison. Compared to the dense tube region, the low-density tube exhibits reduced fluxes at low energies, resulting in a much harder spectrum with a decreased spectral slope. This harder spectrum provides direct evidence of plasma energization by magnetic reconnection.

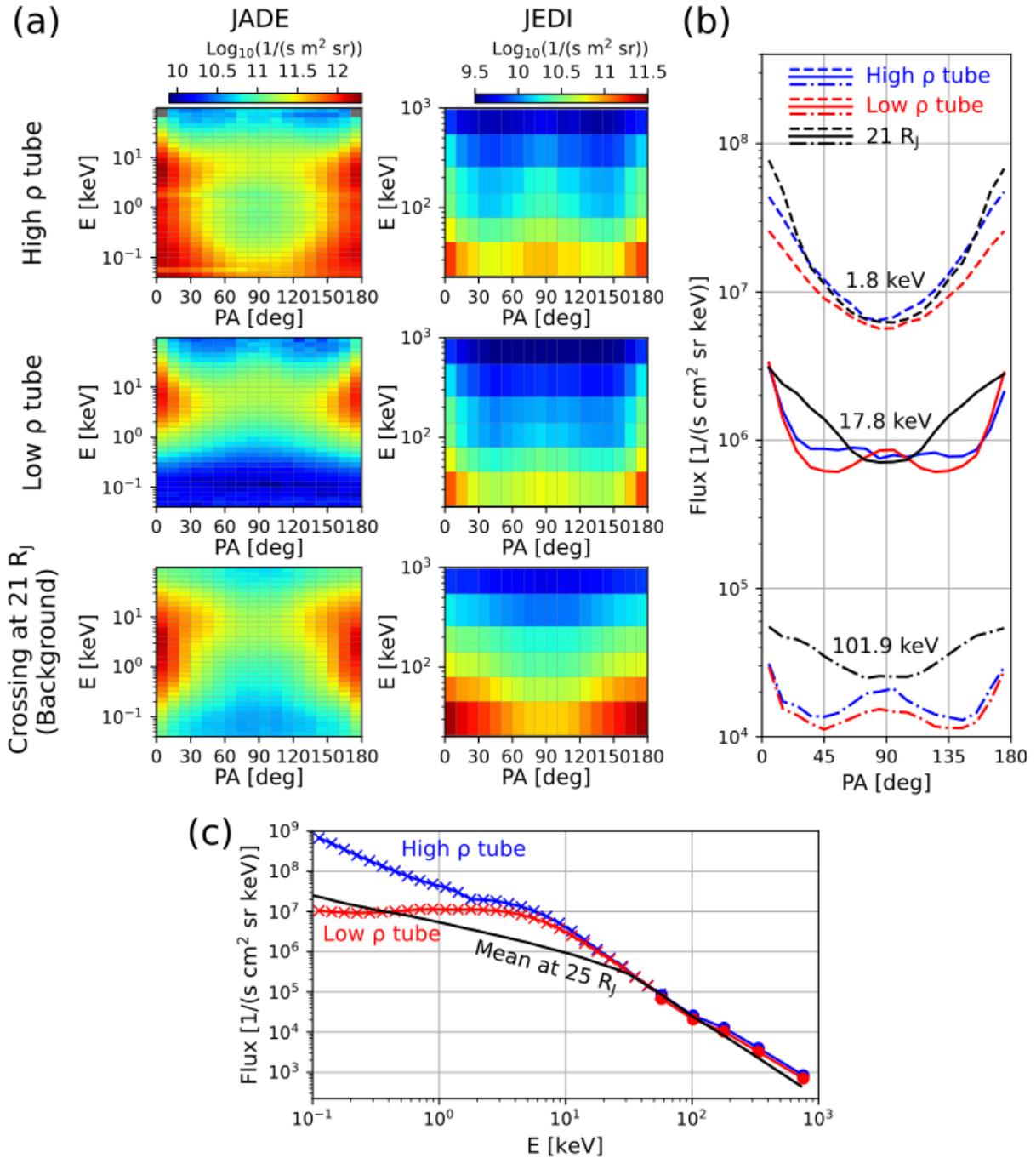

**Figure 3.** (**a,b**) Pitch angle (PA) distributions of electrons in the high-density tube region (02:00:42-02:02:42 UT) and low-density tube region (02:03:45-02:05:45 UT) of the magnetic reconnection event, along with a typical case at the plasma disk center at 21 $R_J$ (05:55:00-06:03:00 UT). (**c**) Electron spectra in the high- and low-density tubes, compared with the mean spectrum at 25 $R_J$ from Liu et al. (2024). Thermal and energetic electrons are detected by the JADE and JEDI instruments, respectively.

**Summary**

Combining the southern entangled flux tubes and the northern current sheet with plasma jets, this study provides the first observational evidence of magnetic reconnection driven by localized flux tube interchange in Jupiter's magnetosphere. As illustrated in **Figure 4**, at $t_1$, the system becomes unstable under centrifugal force due to the configuration of an inner, mass-loaded flux tube and an outer, lower-mass flux tube. The inner tube, experiencing a larger centrifugal force, tends to exchange positions with the outer tube to achieve equilibrium. However, the exchange of two whole flux tubes is constrained by the requirement for the convection of their footprints on the ionosphere. Instead, the equatorial region is more unstable to centrifugal instability, leading to a localized interchange. During this process, the inner partial tube moves outward while the outer partial tube moves inward, causing the two tubes to stretch. Simultaneously, flow convection occurs near the equator, twisting the magnetic field lines locally. Consequently, a pair of off-equator structures with entangled flux tubes forms at $t_2$, with Juno traversing the southern one during this period.

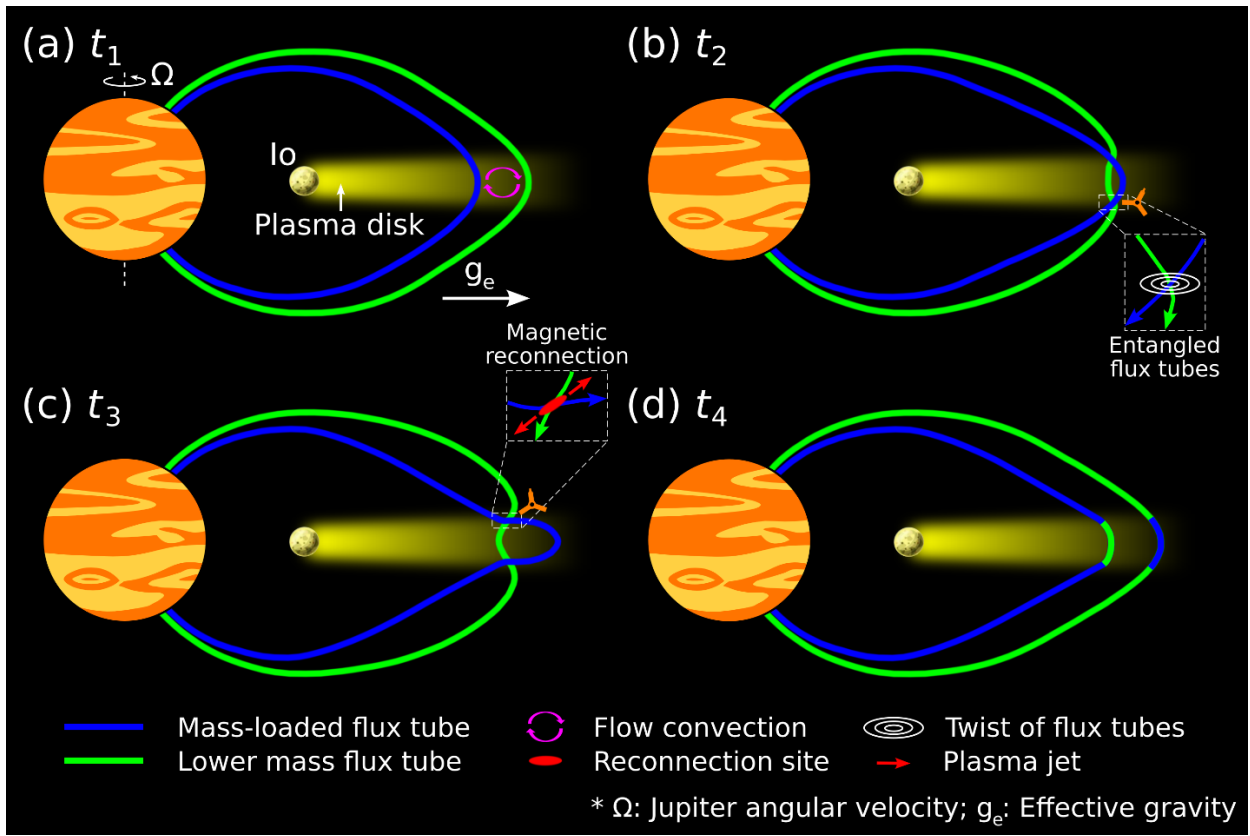

**Figure 4.** Simplified illustration of magnetic reconnection from localized flux tube interchange (not to scale). (**a**) The initial state at $t_1$. (**b**) Twisting of magnetic field lines with the formation of a pair of structures with entangled flux tubes at $t_2$. (**c**) Formation of a pair of reconnection sites at $t_3$. (**d**) Completion of the localized flux tube interchange at $t_4$. The effective gravity ($g_e$) is dominated by the centrifugal force.

As the twisting and stretching motions intensify, magnetic reconnection is triggered at $t_3$ when the shear angle between the magnetic fields of the two flux tubes becomes sufficiently large.

This results in the formation of a pair of off-equator reconnection sites, with Juno traversing the northern site at that time. Two mechanisms could explain the observed vertical plasma jets. First, the magnetic field in the inner tube undergoes stronger stretching due to the larger centrifugal force, causing the 'X-shape' within the dashed-line box of panel (c) to skew and produce northward/southward plasma jets. Second, during the asymmetric reconnection event, the parallel components of the magnetic fields in the two tubes act as a guide field. The outflow tends to align with this guide field, which is vertically oriented in the diagram. Following the reconnection, the two flux tubes merge, completing the localized interchange process at $t_4$. As a result, plasma originating from Io is transported outward, accompanied by the relaxation of the twisted flux tubes.

**Discussion and Conclusions**

In this study, we report a guide field magnetic reconnection event occurring at 23.3 Jupiter radii within the plasma disk, which is the innermost reconnection event observed in Jupiter's magnetosphere. This event is also the first confirmed observation of an ion diffusion region resulting from magnetic reconnection at Jupiter, identified by the unambiguous signature of the asymmetric Hall magnetic field. Ion plasma jets with increased bulk velocity perpendicular to the equator are directly observed through thermal plasma measurements, with the jet offset toward the low-density side due to significant density asymmetry. Additionally, the electron pitch angle distribution reveals plasma energization, characterized by a flux enhancement near 90 deg pitch angles and a harder spectrum after reconnection. Further verification is provided by examining the Walén relation, energy conservation, and the ion inertial scale, all of which show consistency with the observations.

This event is also the first observational evidence of flux tube interchange driven reconnection at Jupiter, supported by the location of the event deep within the magnetosphere, the observation of a pair of structures symmetrically located relative to the plasma disk equator with opposite magnetic field polarities, and the presence of vertically directed plasma jets. It demonstrates that interchange motion involves a complex build-up of twisted flux tubes, which are ultimately relaxed by reconnection. A thin current sheet forms as a result of the collision between two flux tubes with distinct properties, driven by centrifugal force induced stretching and twisting. In this thin current sheet, asymmetric reconnection occurs, characterized by a large guide field and a significant density difference. Moreover, other mechanisms could be ruled out. At 23.3 Jupiter radii, the plasma disk is generally expected to corotate with Jupiter with little perturbation. This much closer location, compared to the Galileo X-line, helps exclude reconnection driven by the Vasyliūnas cycle. It is also far from any Galilean moons, eliminating the influence of moon-plasma interactions. The observed vertically directed jets further allow us to rule out plasma injections.

In the study of Jupiter's magnetosphere, two of the most important open questions are how magnetic reconnection governs the system and how Io-genic plasma is transported outward and ultimately lost. This study sheds new light on both issues with direct evidence and helps refine models of Jovian magnetospheric dynamics. On the one hand, the proposed mechanism of localized flux tube interchange could play a crucial role in making the transport of Io-genic plasma more efficient. On the other hand, this study reveals key characteristics of interchange-driven magnetic reconnection including the complex nature of twisted flux tubes merging, a new reconnection scenario that previously lacked observational evidence. Magnetic reconnection is a key process in plasma physics, and these results offer valuable constraints for future simulations. Finally, this research offers a new perspective on how magnetized plasmas behave in astrophysical systems with rapid rotation, strong magnetic fields, and massive plasma sources, e.g., giant planets, pulsars, brown dwarfs, and accretion disks. Since centrifugally driven localized processes could also occur in these systems, this study highlights its broader astrophysical significance.

## Methods
### Forward Modeling of the Thermal Plasma Data
Using data from the Juno/JADE ion detector (JADE-I), we employ a two-step forward modeling method to derive plasma parameters. Within one spacecraft spin, JADE-I measures ion count rates at different energy levels from multiple views and directions, covering $6\pi$ steradians in High Rate Science mode and $4\pi$ steradians in Low Rate Science mode (i.e., the SPECIES data). JADE-I also records accumulated count rates in different time-of-flight bands (i.e., the TOF data). In the first step, assuming the plasma follows the kappa distribution, we fit the TOF data to determine the abundances of five major ion species: $O^+$, $O^{++}$, $S^+$, $S^{++}$, and $S^{+++}$. In the TOF data, signals from $O^+$ and $S^{++}$ overlap due to their similar mass-to-charge ratios. The density ratio between them is fixed as $n(O^+):n(S^{++}) = 1.2$, inferred from a physical chemistry model in Delamere et al. (2005). In the second step, with fixed ion abundances, we fit the SPECIES data to determine the plasma bulk parameters, including density ($n$), temperature ($T$), and 3-D bulk velocities ($\vec{v}$) in the Jupiter-centered Jupiter-De-Spun-Sun (JSS) coordinate system. In the JSS system, the $z$-axis aligns with Jupiter's spin axis. The $y$-axis is defined as the cross product of $+z$ with the Jupiter-to-Sun vector. The $x$-axis completes the right-handed system and is approximately directed toward the Sun. Details of this forward modeling technique are elaborated in J. Wang et al. (2024a, 2024b).

### Pitch Angle Distributions of Ions and Electrons
The pitch angle $\alpha$ is defined as the angle between a particle's velocity vector and the local magnetic field. For electrons, $\alpha$ values are provided in the JADE-E and JEDI-E datasets and are used directly. In contrast, ion pitch angles require additional calculation. The JADE-I dataset provides the local magnetic field vector and the look direction ($\hat{\theta}_l$) for each measurement, allowing $\alpha$ to be computed as: $\alpha = \arccos(-\hat{\theta}_l \cdot \vec{B}/|\vec{B}|)$. For low-energy ions, pitch angle distributions are influenced by plasma bulk velocity (Sarkango et al., 2023). To minimize this bias, we restrict ion analysis to energies above 10 keV/q, where ion kinetic velocities significantly exceed bulk flow velocity. Once $\alpha$ is determined for each data point, the data are binned to obtain the pitch angle dependent flux spectra ($J'$) by averaging over all data points ($J$):

$$J'(\alpha_m, E, t) = \frac{\sum_i J(\alpha_i, E, t)[\alpha_m - \Delta\alpha/2 < \alpha_i < \alpha_m + \Delta\alpha/2]}{\sum_i [\alpha_m - \Delta\alpha/2 < \alpha_i < \alpha_m + \Delta\alpha/2]}$$

where $[\alpha_m - \Delta\alpha/2, \alpha_m + \Delta\alpha/2]$ defines the $m$-th pitch angle bin with width $\Delta\alpha$. $E$ and $t$ denote energy level and time, respectively. The term $[\alpha_m - \Delta\alpha/2 < \alpha_i < \alpha_m + \Delta\alpha/2]$ equals 1 if true and 0 if false.

**Determination of the LMN Coordinate System**
In the analysis of the ion diffusion region, we follow a similar approach to Eriksson et al. (2024) to establish the local current sheet LMN coordinate system. Due to the limitations of single-spacecraft observations, the LMN system based on the cross-product normal is more robust than using the three eigenvectors of the Minimum Variance Analysis (MVA) of the magnetic field data, as illustrated in R. Wang et al. (2024) and Knetter et al. (2004). The boundary-normal direction, $\vec{N}$, is defined as $\vec{B}_1 \times \vec{B}_2 / |\vec{B}_1 \times \vec{B}_2|$, where $\vec{B}_1$ and $\vec{B}_2$ are magnetic field measurements adjacent to the candidate current sheet. In the reconnection event, the left boundary of the current sheet was identified at 02:02:56 UT, with $\vec{B}_1$ as the average magnetic field from three data points with a resolution of 1 s centered around this time. Similarly, $\vec{B}_2$ is the magnetic field observation at the right boundary, identified at 02:03:23 UT. The out-of-plane (or guide magnetic field) direction is $\vec{M} = \vec{N} \times \vec{L}_{MVA} / |\vec{N} \times \vec{L}_{MVA}|$, where $\vec{L}_{MVA}$ is the direction of the maximum magnetic field variance determined by MVA over the interval of current sheet observations (i.e., from 02:02:56 UT to 02:03:23 UT). Finally, the reconnection exhaust direction is obtained by completing the orthogonal system: $\vec{L} = \vec{M} \times \vec{N}$. In the MVA, the three eigenvalues are $\lambda = $ (458.14, 7.18, 0.66), and the large ratios between them confirm that $\vec{L}_{MVA}$ is well-defined. As a result, the determined vectors in the JSS coordinate system are organized into $r$, $\theta$, and $\phi$ components: $\vec{L} = $ (-0.11, -0.85, 0.51), $\vec{M} = $ (0.22, 0.48, 0.85), and $\vec{N} = $ (-0.97, 0.20, 0.14). We compared $\vec{B}_1$ and $\vec{B}_2$ at different times relative to the current sheet, which always produce similar $\vec{M}$ and show similar Hall magnetic field patterns, which are crucial for identifying the diffusion region.

**Estimations of Key Parameters in the Magnetic Reconnection Event**
Several key parameters are calculated to confirm the detection of the magnetic reconnection event and the diffusion region. First, to check the Walén relation, we calculate the Alfvén speed as $v_A = B/\sqrt{\mu_0 n m_i} = 91$ km/s, where $\mu_0$ is the vacuum permeability, $B = 41$ nT is the average magnetic field intensity, $n = 4$ cm$^{-3}$ is the plasma density in the high density region, and $m_i = 24$ amu is the average ion mass in the plasma disk. To check the energy conservation law, the input magnetic energy density is calculated as $E_b = B^2/(2\mu_0) = 6.7 \times 10^{-10}$ J/m$^3$. The gained kinetic energy density of the outflowing plasma is $E_k = 0.5 n m_i \Delta v^2 = 8.0 \times 10^{-10}$ J/m$^3$, where $\Delta v = 100$ km/s is the increase in plasma flow velocity. To check the diffusion region scale, the ion inertial length is calculated as $d_i = C/\omega_{pi} = 370$ km, where $C$ is the speed of light. $\omega_{pi} = \sqrt{nZ^2 e^2/(\epsilon_0 m_i)}$ is the ion plasma frequency, where $\epsilon_0$ is the permittivity of free space and $e$ is the elementary charge. $Z = 1.5$ is the average charge state of ion plasma in the plasma disk.

**Extended Data**

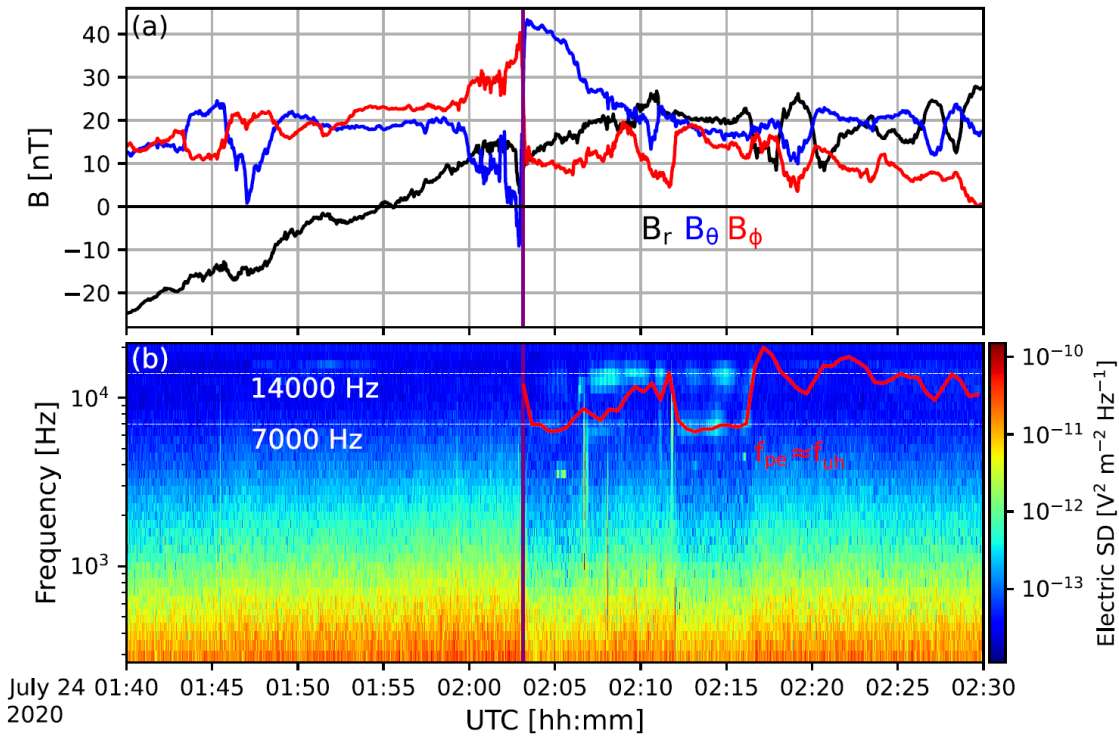

**Extended Data Figure 1. Plasma wave observations associated with the magnetic reconnection event.** Panels (a) and (b) show the magnetic field in the JSS coordinate system measured by Juno's MAG instrument and the spectral density (SD) measured by the electric antenna of Juno's WAVES instrument, respectively. Harmonic plasma waves near 7,000 Hz and 14,000 Hz are observed. The observed wave frequencies are comparable to the electron plasma frequency ($f_{pe}$) and the upper hybrid frequency ($f_{uh}$), as indicated by the red solid line. The thick, solid purple line indicates the location of the current sheet.

**Acknowledgements**
We thank the Juno JADE team, without whose support this research would not be possible. This work was supported at the University of Colorado as a part of NASA's Juno mission funded by NASA through contract 699050X with the Southwest Research Institute.

**Competing Interests**
The authors declare no competing interests.

**Data Availability**
All the Juno observations used in this research are publicly available from the NASA Planetary Data System Plasma Interactions Node. The Juno JADE data are the level 3 version 04 data from https://doi.org/10.17189/2775-4623. Electron density and temperature are Juno JADE derived moments (level 5 version 01) from https://doi.org/10.17189/2fch-6v84. The Juno magnetometer data are the level 3 version 01 data from https://doi.org/10.17189/1519711. The Juno JEDI data are the level 3 version 01 data from https://doi.org/10.17189/1519713. The plasma bulk parameters are available at https://doi.org/10.5281/zenodo.12802043.

## Author Contributions

Supervised by F.B., J.-Z.W. wrote the manuscript and performed the data analysis. S.E., R.E.E., F.B., and L.C.R. contributed to the interpretation of the reconnection event. S.E. contributed to the current sheet coordinate analysis. P.A.D. contributed to the flux tube interchange driven reconnection model. R.J.W. contributed to producing the JADE data. R.W.E., P.W.V., and F.A. contributed to building the JADE instrument, on which the study is based. R.J.W. and R.W.E. contributed to the JADE data analysis. All authors contributed to discussing the results and editing the manuscript.